\documentclass[twocolumn,prb]{revtex4}

\usepackage{graphicx,amsmath}

\begin{document}

\title{Superfluid-insulator transition and BCS-BEC crossover in "dirty" ultracold Fermi gas}

\author{B. I. Shklovskii}

\affiliation{Theoretical Physics Institute, University of
Minnesota, Minneapolis, Minnesota 55455}

\date{\today}

\begin{abstract}

Superfluid-insulator transition in an ultracold Fermi gas in the
external disorder potential of the amplitude $V_0$ is studied as a
function of the concentration of the gas $n$ and magnetic field
$B$ in the presence of the Feshbach resonance.  We find the zero
temperature phase diagrams in the plane ($B,n$) at a given $V_0$
and in the plane $(V_0, n)$ at a given $B$. Our results for BEC
side of the diagram are also valid for the superfluid-insulator
transition in a Bose gas.

\end{abstract}

\maketitle

Using the Feshbach resonance in the magnetic field $B$ one can
study lots of interesting physics in ultracold Fermi gases (see
the recent review article~\cite{Pit} and references therein). In
the vicinity of the Feshbach resonance $B=B_0$ the scattering
length of two fermions typically changes as
\begin{equation}
a= a_{0}\frac{\Delta B}{B_0 - B} , \label{a}
\end{equation}
where we omitted the non-resonant term. As a result by the
decreasing magnetic field the Fermi gas can be transformed from
the phase of weakly attracting fermions (at $B > B_0$) to the
phase of repelling each other compact composite bosons, dimers
made of two fermions with opposite spins (at $B < B_0$). At $B =
B_0$ the gas goes through unitarity~\cite{Pit}. In a clean Fermi
gas all mentioned above phases are superfluid. Far enough from the
resonance at $B > B_0$ superfluidity is described by the
Bardeen-Cooper-Schrieffer (BCS) theory, while on the other side,
at $B < B_0$, the theory of Bose-Einstein condensation (BEC) of
composite bosons works. Thus, reduction of magnetic field $B$
leads the gas through BCS-BEC crossover.

The aim of this paper is to consider the zero temperature BCS-BEC
crossover in a "dirty" Fermi gas, i.e. in the gas situated in a
three-dimensional (3D) random potential. Such a random potential
can be created, for example, by superposing a 3D speckle on the
ultracold gas sitting in a trap. Obviously, a strong enough
disorder can localize the Fermi gas on BCS side and the BEC
condensate on BEC side, destroying superfluidity in both cases.
For brevity, we call the localized phase "insulator" and the
localization transition "superfluid-insulator" (SI) transition. In
this paper we are talking about SI transition in uniform in
average infinite gas but some of our results can be applicable to
experiments with wide enough traps.

Expansion of BEC condensate of ultracold Bose gases in the
disorder potential of one-dimensional speckles has been recently
studied experimentally~\cite{Fort,Clement}. It was found that the
disorder stops expansion at some distance. In this case, however,
a big role may be played by rare very high hills of the random
potential. Apparently several laboratories are planning similar
studies of SI transition in a potential created by 3D speckles.
One can expect that in this case the rare high hills are less
important and theory of SI transition in infinite system is more
relevant.

\begin{figure}[b]
\centerline{\includegraphics[width=2.6in]{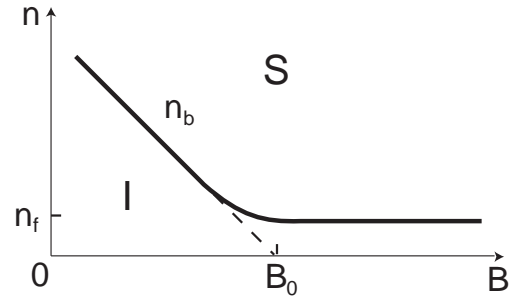}}
\caption{ \label{Fig:Diagram} The phase diagram of the SI
transition. Magnetic field $B$ is plotted on the horizontal axis,
while the fermion concentration $n$ is plotted on the vertical
one. S stands for superfluid and I for insulator. $B_0$ is the
Feshbach resonance point. Critical concentrations $n_f$ and
$n_{b}(B)$ for the classical random potential are straight lines
given by Eqs.~(\ref{n1}) and (\ref{n2}) respectively.}
\end{figure}

In a Fermi gas in a fixed external random potential the SI
transition can be driven by the decreasing concentration of
fermions $n$ at a given magnetic field $B$, or by the decreasing
$B$ at a given $n$. Therefore, one can think about the SI phase
diagram of a Fermi gas in the plane ($B,n$). In this paper, we
find the zero temperature SI border line $n(B)$ on such a phase
diagram (see Fig.~\ref{Fig:Diagram}). Because $B$ and $n$ can be
independently controlled this diagram can be verified
experimentally.
We characterize disorder by the amplitude of the random potential
energy $V_0$ (mean square deviation of random potential $V(r)$
from average value) and the characteristic size of potential wells
and hills $R$. In the first part of this paper we assume that both
$V_0$ and $R$ are so large that if $m$ is the mass of the fermion
\begin{equation}
V_0 \gg \frac {\hbar^2}{mR^{2}} . \label{V}
\end{equation}
This means that a typical potential well has many levels. In this
sense we are talking about classical random potential. From the
beginning we assume that $n(B)$ is so large that the average
number of atoms in a well $nR^3 \gg 1$, but in the end we show
that near the SI border this inequality follows from (\ref{V}).

Let us first consider BCS phase corresponding to $B > B_0$. Here
the criterion of superfluidity coincides with the condition that
the Fermi level of weakly interacting Fermi gas reaches the
mobility edge in a given external potential. Roughly speaking,
this happens when the Fermi energy of the gas $E_F =
(\hbar^2/2m)(3\pi^2 n)^{2/3}$ becomes larger than the amplitude
$V_0$ of the random potential. This condition leads to the
critical concentration of SI transition $n=n_f$ on the side of
free fermions $(B > B_0)$
\begin{equation}
n_f = C_{f} R^{-3}\left ( \frac{V_0}{\hbar^{2}/m R^2}
\right)^{3/2} . \label{n1}
\end{equation}
Thus, the segment of the SI border at $B > B_0$  is horizontal as
shown in Fig.~\ref{Fig:Diagram}. It is known that in a classical
long range  potential the numerical coefficient $C_f$ can be found
using the idea that in the classical potential (\ref{V}) the
mobility edge coincides with the classical percolation level
$V_{p}$ ~\cite{Zallen,book}. This is the level at which Fermi gas
lakes formed in the random potential wells merge to create the
infinite cluster or the Fermi sea. In a generic three-dimensional
gaussian potential with the distribution function
\begin{equation}
F(V)= \frac{1}{V_{0}\sqrt{2\pi}} \exp(-V^{2}/2V_{0}^{2}) ,
\label{Gauss}
\end{equation}
this level corresponds to occupation of $\theta_c = 17 \%$ of the
space by lakes~\cite{Skal,book}. This gives
\begin{equation}
V_{p} = - 0.96 V_{0}. \label{perc}
\end{equation}
Now we can find  $n_f$ as the total concentration of fermions in
wells deeper than $V_{p}$. Inequality~(\ref{V}) lets us use
Thomas-Fermi (TF) approximation
\begin{equation}
n_{f} = \frac{1}{3\pi^2} \int_{-\infty}^{V_p} \left(
\frac{2m(V_{p}-V)}{\hbar^2}\right)^{3/2} F(V)dV  . \label{nlarger}
\end{equation}
For a gaussian potential this leads to the coefficient in
Eq.~(\ref{n1})
\begin{equation}
C_f = \frac{2}{3\pi^{5/2}} \int_{-\infty}^{V_{p}/V_0} [V_p/V_0 - x
]^{3/2}\exp(-x^2/2)dx \simeq 0.008. \label{clarger}
\end{equation}
Let us switch to the less trivial BEC side of the diagram which
corresponds to $B < B_0$. In this case, interaction of dimers
plays the crucial role. Following Ref.~\cite{Pit} we refer to the
scattering length of the two dimers as $a_{dd}$. Then the uniform
gas of interacting dimers has the positive chemical potential
\begin{equation}
\mu(n) = \frac {4\pi \hbar^{2} (n/2) a_{dd}}{2m} = \frac {\pi
\hbar^{2} n a_{dd}}{m} . \label{mu}
\end{equation}
Here we took into account that the concentration of dimers is
$n/2$, while the dimer mass is $2m$. If $\mu(n)$ is larger than
the amplitude of the random potential, $V_0$, the gas of dimers
can screen the random potential redistributing a small fraction of
its density from the hills of the random potential to the wells.
On the other hand, if $\mu(n) \ll V_0$ the gas is fragmented in
many disconnected lakes. Thus, the condition of delocalization of
dimers and, therefore, the condition of superfluidity in this case
is roughly speaking $\mu(n) = V_0$. Substituting the nontrivial
result of Ref.~\cite{Shlyap}
\begin{equation}
a_{dd} = 0.6a \label{add}
\end{equation}
into Eq.~(\ref{mu}) and using Eq.~(\ref{a}) we get for the SI
border concentration of fermions $n_b$ on the compact bosons side
( $B < B_0$)
\begin{equation}
n_{b}(B) = C_b R^{-3} \frac {R}{a}\frac{V_0}{\hbar^{2}/mR^2}= C_b
R^{-3} \frac {R}{a_0}\frac{V_0}{(\hbar^{2}/mR^2)}
 \frac{B_{0}-B}{\Delta B}. \label{n2}
\end{equation}
Let us now estimate the numerical coefficient $C_b$. To this end
we again appeal to the percolation theory and deal with the
percolation level $2V_p$ in the potential energy of a dimer
$2V(r)$. The local concentration $n(r)/2$ of dimers adjusts to
external potential according to the Gross-Pitaevskii equation
(GPE)
\begin{equation}
\mu\psi(r) = \left[- \frac{ \hbar^{2}\nabla^2}{4m} + 2V(r) +
\frac{4\pi \hbar^{2} a_{dd}}{2m} |\psi(r)|^{2}\right]\psi(r),
\label{GP}
\end{equation}
where $\mu$ is the condensate chemical potential, $2V(r)$ is the
potential acting on a dimer, and the condensate wave function
$\psi(r)$ is normalized to total number of dimers, $\int dr
|\psi(r)|^{2} = N/2$, where $N$ is the total number of fermions.
Thus, $|\psi(r)|^{2}$ has the meaning of the local concentration
of dimers $n(r)/2$. Let us show that near the SI border one can
use the TF approximation and drop the kinetic energy term of GPE.
This can be done if the healing length $l_h =[(n/2)a_{dd}]^{-1/2}$
of the condensate is much smaller than characteristic length of
potential, $l_h \ll R$. Using above estimate for the critical
concentration $n_b$ we get that for the classical disorder
potential (Eq.~(\ref{V})) at the BEC side SI border
\begin{equation}
\frac{l_h(n_b)}{R} =
\left(\frac{\hbar^{2}/mR^{2}}{V_0}\right)^{1/2} \ll 1. \label{lh}
\end{equation}
Thus, one can proceed in the TF approximation, where at every
point local concentration of the condensate $n(r)$ satisfies
equation
\begin{equation}
\frac{\pi \hbar^{2}n a_{dd}}{m} + 2V(r) = \mu. \label{muglobal}
\end{equation}
The chemical potential $\mu$ is determined by normalization of
concentration of dimers $n(r)/2$ to the total number of dimers
$N/2$ and grows with increasing $N$. If $\mu < 2V_p$ we get only
disconnected Bose gas lakes. If $\mu > 2V_p$ the merging lakes
form the Bose sea or the infinite cluster. Thus, similarly to the
BCS side on the BES side the SI transition also happens when $\mu
= 2V_p$. For a gaussian potential with the help of Eq.~(\ref{add})
this gives for $C_b$ in Eq.~(\ref{n2}).
\begin{equation}
C_b = \frac{\sqrt{2}}{0.6\pi^{3/2}} \int_{-\infty}^{V_{p}/V_0}
[V_p/V_0 - x]\exp(-x^{2}/2)dx \simeq 0.01. \label{csmaller}
\end{equation}
For more realistic~\cite{Lugan} distribution of the speckle
potential we do not know the percolation threshold, but on the
basis of approximate universality~\cite{book} of the $\theta_c$ we
guess that $C_b$ is the same as for gaussian potential within $20
\%$.

Thus, the SI border $n(B)$ consists of the two straight lines, as
shown in Fig.~\ref{Fig:Diagram}. At $B < B_0$ it follows the line
with the negative slope, Eq.~(\ref{n2}) and at $B > B_0$ the
border line is horizontal, Eq.~(\ref{n1}). Eq.~(\ref{n2}) is valid
until $n_{b}(B) \gg n_f$. At $B = B_0 - \delta B$, where $\delta B
= \Delta B [V_0/(\hbar^{2}/m a_{0}^2)]^{1/2}$ Eq.~(\ref{n2})
crosses over to Eq.~(\ref{n1}). In the unitarity interval of the
width $\delta B$ around $B=B_0$ at the SI border we arrive at
$n(B)a^{3}\sim 1$. In other words in this interval $k_{F}a \sim
l_h/a\sim 1$. Because the length of a dimer, $\xi \sim a_{dd} \sim
a$, one can also say that only in the unitarity interval $n\xi^3
\sim 1$ and dimers touch each other. This differs from the
superconductor-insulator transition in a toy Coulomb
model~\cite{Shklovskii}, where the $n\xi^3 \sim 1$ at the long
intermediate segment of the superconductor-insulator border.

The fact that the critical concentration $n_b$  exceeds $n_f$ is
easy to understand. Indeed, at a given $n$ dimers have much
smaller chemical potential than weakly interacting fermions. Thus,
dimers need a larger concentration $n$ in order to get
delocalized. In the similar way one can understand the growth of
$n(B)$ with the decreasing $B$ at $B < B_0 $. Indeed, at a given
$n$, the farther from the resonance, the more ideal the Bose gas
of dimers is, the smaller is its chemical potential. Again, to
compensate for this trend $n(B)$ should grow with the decreasing
$B$.

Above we assumed that the number of particles in a well of the
random potential is large, $nR^{3}>> 1$ and used the mean field
approximation on the BEC side, ignoring discreetness of particles.
As we see from Fig.~\ref{Fig:Diagram} the minimum value of the
border concentration $n(B)$ is $n_f$. Therefore, inequality
$n_{f}R^3 \gg 1$ guarantees that everywhere on the border $n(B)R^3
\gg 1$. Substituting Eq.~(\ref{n1}) into $n_{f}R^3 \gg 1$ we
arrive at inequality (\ref{V}). Thus, this inequality is the
single condition of validity of the above theory of SI border.

It is clear from the above discussion that the insulating phase on
the BEC side consists of disconnected lakes, populated by dimers.
One can use the term Bose glass~\cite{Fisher} for this phase,
because it has no excitation gap and is compressible.

Until now we assumed that the disorder is classical in the sense
of inequality (\ref{V}). Let us now discuss what happens for a
quantum random potential, when the opposite strong inequality
\begin{equation}
V_0 \ll \frac {\hbar^2}{mR^{2}} \label{Vweak}
\end{equation}
takes place. In an experiment one can move from inequality
(\ref{V}) to inequality (\ref{Vweak}) by scaling down the
intensity of the light beams creating speckles, while keeping the
rest of the speckle set up (including $R$) fixed. How will then
the phase diagram in ($B,n$) plane change?
\begin{figure}[b]
\centerline{\includegraphics[width=2.0in]{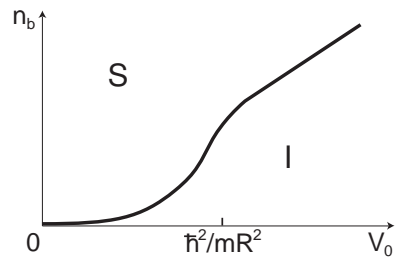}}
\caption{ \label{Fig:Diagram1} Schematic phase diagram of SI
transition for a Fermi gas at $B < B_0$ or for a weakly non-ideal
Bose gas. The critical concentration of SI transition $n_b$ is
plotted against the disorder amplitude $V_0$. S stands for
superfluid and I for the insulator. The SI border shows crossover
between regimes of quantum and classical random potentials at $V_0
= \hbar^{2}/mR^{2}$.}
\end{figure}

Let us start this discussion from the BCS side of the diagram
($B>B_0$) and concentrate on the disorder induced density of
states (DOS) at small energies. For simplicity, we assume that we
are dealing with a gaussian potential $V(r)$ which two point
correlation function decays as $1/r^{3}$ or faster at $r \gg R$.
According to inequality (\ref{Vweak}) the case of quantum disorder
the wells of the size $R$ do not have levels. In this case, the
characteristic energy of the low energy tail of DOS is determined
by wells of the size $L \gg R$, which are large enough to get a
level~\cite{Halperin,Baranovskii,book}. The typical depth of such
wells $V(L)$ is much smaller than $V_0$, namely $V(L)=
V_{0}(R/L)^{3/2}\ll V_0$. This happens due to the cancellation of
the majority of $(L/R)^{3}$ contributions of wells the and hills
of the size $R$. Using condition of the level existence $V(L) =
\hbar^2/mL^{2}$, we find the characteristic size of the well,
which has a single level
\begin{equation}
L_c = R \left(\frac{\hbar^2}{mR^{2}V_0}\right)^{2} . \label{L}
\end{equation}
Substituting $L_c$ to $V(L)$ we arrive at the characteristic
energy scale of the low energy
tail~\cite{Halperin,Baranovskii,book}

\begin{equation}
V_t = C_{t}V_0 \left( \frac {V_0}{\hbar^{2}/mR^{2}}\right)^{3} .
\label{VT}
\end{equation}
The energy which separates localized and delocalized states (the
mobility edge) is also of the order of $V_t$. For a quantum random
potential (Eq.~(\ref{Vweak})) $V_t \ll V_0$ and the concentration
of fermions which can be localized in the tails or, in other
words, the critical concentration of SI transition, $n_{f}$ is
very small, too
\begin{equation}
n_{f} \sim R^{-3}\left(\frac
{V_0}{\hbar^{2}/mR^{2}}\right)^{6},~~~~(V_0\ll \hbar^{2}/mR^{2}).
\label{nlargerstar}
\end{equation}
Let us switch now to the BEC side of the phase diagram ($B <B_0$).
In this case, the tails of DOS can accommodate more dimers in the
band of energies $V_t$ because we can condense many bosons at one
level. Only if the chemical potential of bosons given by
Eq.~(\ref{mu}) becomes larger than $V_t$ the states become
delocalized. Thus, $n_{b}$ can be estimated equating $\mu$ and
$V_t$. Using Eq.~(\ref{mu}) and Eq.~(\ref{VT}) we get for $V_0\ll
\hbar^{2}/mR^{2}$
\begin{equation}
n_{b} \sim R^{-3}\frac{R}{a} \left(\frac
{V_0}{\hbar^{2}/mR^{2}}\right)^{4} \sim  R^{-3}\frac{R}{a_0}
\left(\frac
{V_0}{\hbar^{2}/mR^{2}}\right)^{4}\frac{B_{0}-B}{\Delta B}.
\label{nsmallerstar}
\end{equation}
This results can be also obtained from the
condition~\cite{Shapiro} that the SI transition happens at $\mu
\tau /\hbar \sim 1$, where $\tau$ is the relaxation time of a
delocalized boson with the energy $\mu$. The crossover between
Eqs. (\ref{nlargerstar}) and (\ref{nsmallerstar}) happens in the
unitarity interval of the width $\delta B = \Delta B (a_0/R)
[V_0/(\hbar^{2}/mR^{2})]^{2}$ around $B_0$. In this interval
$na^{3}\sim k_{F}a \sim l_h/a\sim L_c/a \sim 1$ similarly to the
case of the classical potential. In the language of GPE the
estimates we arrived above correspond to the solution, where all
three terms in the right side of Eq.~(\ref{GP}) play comparable
roles. In other words, expectations of the kinetic energy term, of
the random potential term and of the repulsion energy are of the
same order of magnitude at SI border. Note, that at the same time
the amplitude $V_0$ of the bare potential is much larger than
other terms. Only the quantum mechanical averaging makes the
disorder potential energy $ V(L_c)$ equal to other terms. For a
Fermi gas, the idea of such averaging is known, for a long
time~\cite{Halperin,Baranovskii,book}. For a weakly non-ideal Bose
gas idea of quantum mechanical "smoothing" of disorder potential
was explored only recently~\cite{Sanchez,Lugan}. However, SI phase
diagram of an infinite, uniform in average gas could not be
studied in Refs.~\cite{Sanchez,Lugan} because they dealt with an
one-dimensional disorder potential.

Let us discuss applicability of the mean field theory (GPE) for
calculation of $n_b$. GPE is applicable if at $n = n_b$ the
characteristic length $L_c$ or if $n_b L_{c}^{3} \gg 1$. It is
clear that $n_b \gg  n_f$. Multiplying this inequality by
$L_{c}^{3}$ and using Eqs.~(\ref{L}) and (\ref{nsmallerstar}) we
arrive at necessary inequality $n_b L_c^3 \gg n_f L_c^3 \sim 1$.
Thus, the mean field theory is applicable for calculation of
$n_b$. Mean field approach fails and one arrives at
single-particle regime~\cite{Lugan} only at $ n\ll n_b$. We see
from  Eqs. (\ref{nlargerstar}) and (\ref{nsmallerstar}) that in
the case of a quantum random potential  $V_{0}\ll
\hbar^{2}/mR^{2}$ both critical concentrations $n_{f}$ and $n_{b}$
decrease very rapidly with the decreasing $V_0$. As a result,
while the whole phase diagram in this case looks like
Fig.~\ref{Fig:Diagram}, the concentrations $n_{f}$ and $n_{b}$ are
dramatically smaller than for a classical random potential.

The summary of our results for the BEC phase is given in
Fig.~(\ref{Fig:Diagram1}) where we plot the critical concentration
of SI transition $n_b$ as a function of the amplitude of the
random potential $V_0$ (or the intensity of the speckle-building
light), while scattering length $a$ and the characteristic scale
of disorder $R$ are fixed. The fourth order parabola in the
beginning of $n_{b}(V_0)$ curve is given by
Eq.~(\ref{nsmallerstar}). At $V_0 = \hbar^2/mR^2$ this parabola
crosses over to Eq.~(\ref{n2}). Our results for the BEC phase of
dimers shown on Fig.~(\ref{Fig:Diagram1}) are clearly applicable
to a generic weakly non-ideal Bose gas with the scattering
amplitude $a > 0 $.

In conclusion I have studied the zero temperature phase diagram of
the superfluid-insulator phase transition for a Fermi gas going
through Feshbach resonance with the changing magnetic field and
for a Bose gas. I dealt with uniform infinite gases, did not
consider the role of the inverse parabolic profile $n(r)$ in the
trap, and did not study dynamics of the BEC phase expansion when
the trap is eliminated. A likely scenario of this expansion
(similar to that of Ref.~\cite{ClementTheory,Shapiro}) is as
follows. If the maximum concentration of the gas in the center of
the trap $n_m < n_b$ there is no expansion. On the other hand, if
$n_m > n_b$ only bosons from the central domain $r < r_0$ where
$n(r)> n_b$ leave reducing original $n(r)$ to the flat $n (r) =
n_b$ at $r < r_0$. This means that  at $n_m - n_b \ll n_b$ the
fraction of the gas mass released in an expansion is proportional
to $(n_m - n_b)^{3/2}$.

This paper is dedicated to the memory of V. I. Perel. I am indebt
to N. Cooper, L.~Sanchez-Palencia, and G. Shlyapnikov for critical
reading of the manuscript and useful advices. I am grateful to Tao
Hu, P. Lee, S. Sachdev, J. Schmalian, B. Shapiro, D. Sheehy, Z.
Tesanovic for useful discussions. I acknowledge hospitality of the
Aspen Center for Physics, where the idea of this paper was
conceived.


\end{document}